\documentclass[%
reprint,
superscriptaddress,
showkeys,
amsmath,amssymb,
aps,
prb,
]{revtex4-1}
\usepackage{graphicx}
\pdfoutput=1
\usepackage{placeins}
\usepackage{comment}
\usepackage{amsmath}
\usepackage[caption=false]{subfig}
\usepackage{framed}
\usepackage[normalem]{ulem}
\usepackage[export]{adjustbox}
\usepackage{dcolumn}
\usepackage{bm}
\usepackage{array,ragged2e}
\usepackage{booktabs}
\usepackage{lettrine}
\newcolumntype{P}[1]{>{\Centering\hspace{0pt}}p{#1}}
\begin{document}
\preprint{APS/123-QED}
\title[Defect Engineering and Substrate Effects in Ion-Irradiated Monolayer 2D Materials]{Defect Sizing, Separation and Substrate Effects in Ion-Irradiated Monolayer 2D Materials}
\author{Pierce Maguire}
\affiliation{School of Physics, Trinity College Dublin, Dublin 2, Ireland}
\affiliation{AMBER Centre, CRANN Institute, Trinity College Dublin, Dublin 2, Ireland}

\author{Daniel S. Fox}
\affiliation{School of Physics, Trinity College Dublin, Dublin 2, Ireland}
\affiliation{AMBER Centre, CRANN Institute, Trinity College Dublin, Dublin 2, Ireland}

\author{Yangbo Zhou}
\affiliation{School of Physics, Trinity College Dublin, Dublin 2, Ireland}
\affiliation{AMBER Centre, CRANN Institute, Trinity College Dublin, Dublin 2, Ireland}
\affiliation{School of Material Science and Engineering, Nanchang University, Youxun W Rd, Xinjian Qu, Nanchang Shi, Jiangxi Sheng, People's Republic of China}

\author{Qianjin Wang}
\affiliation{National Laboratory of Solid State Microstructures, Nanjing University, Nanjing 210093, Jiangsu Province, People's Republic of China}

\author{Maria O'Brien}
\affiliation{AMBER Centre, CRANN Institute, Trinity College Dublin, Dublin 2, Ireland}
\affiliation{School of Chemistry, Trinity College Dublin, Dublin 2, Ireland}

\author{Jakub~Jadwiszczak}
\affiliation{School of Physics, Trinity College Dublin, Dublin 2, Ireland}
\affiliation{AMBER Centre, CRANN Institute, Trinity College Dublin, Dublin 2, Ireland}

\author{Conor P. Cullen}
\affiliation{AMBER Centre, CRANN Institute, Trinity College Dublin, Dublin 2, Ireland}
\affiliation{School of Chemistry, Trinity College Dublin, Dublin 2, Ireland}

\author{John McManus}
\affiliation{AMBER Centre, CRANN Institute, Trinity College Dublin, Dublin 2, Ireland}
\affiliation{School of Chemistry, Trinity College Dublin, Dublin 2, Ireland}

\author{Niall McEvoy}
\affiliation{AMBER Centre, CRANN Institute, Trinity College Dublin, Dublin 2, Ireland}
\affiliation{School of Chemistry, Trinity College Dublin, Dublin 2, Ireland}

\author{Georg S. Duesberg}
\affiliation{AMBER Centre, CRANN Institute, Trinity College Dublin, Dublin 2, Ireland}
\affiliation{School of Chemistry, Trinity College Dublin, Dublin 2, Ireland}
\affiliation{Institute of Physics, EIT 2, Faculty of Electrical Engineering and Information Technology, Universit{\"a}t der Bundeswehr M{\"u}nchen, Werner-Heisenberg-Weg 39, 85577 Neubiberg, Germany}

\author{Hongzhou Zhang}
\email{Corresponding author E-mail: Hongzhou.Zhang@tcd.ie}
\affiliation{School of Physics, Trinity College Dublin, Dublin 2, Ireland}
\affiliation{AMBER Centre, CRANN Institute, Trinity College Dublin, Dublin 2, Ireland}
\date{\today}
\begin{abstract}
Precise and scalable defect engineering of 2D nanomaterials is acutely sought-after in contemporary materials science.
Here we present defect engineering in monolayer graphene and molybdenum disulfide (MoS$_2$) by irradiation with noble gas ions at 30 keV. Two ion species of different masses were used in a gas field ion source microscope: helium (He$^+$) and neon (Ne$^+$). A detailed study of the introduced defect sizes and resulting inter-defect distance with escalating ion dose was performed using Raman spectroscopy. Expanding on existing models, we found that the average defect size is considerably smaller for supported than freestanding graphene and that the rate of defect production is larger. We conclude that secondary atoms from the substrate play a significant role in defect production, creating smaller defects relative to those created by the primary ion beam.
Furthermore, a similar model was also applied to supported MoS$_2$, another promising member of the 2D material family. Defect yields for both ions were obtained for MoS$_2$, demonstrating their different interaction with the material and facilitating comparison with other irradiation conditions in the literature.
\end{abstract}
\maketitle
\section{\label{sec:level1}Introduction}

\lettrine[lines=2,lhang=0]{I}{n} recent years, the extraordinary properties and tunability of 2D materials have been repeatedly demonstrated, heralding a new era of materials science \cite{Novoselov2004,Novoselov2012}. Their physical properties (electrical, thermal, etc.) are highly distinguished from their bulk counterparts due to the evolution of band structure with decreasing layer number \cite{Butler2013, balleste2010, Backes2014, Splendiani2010}. The ideas and methodologies developed from the investigation of graphene have been extended to many other 2D materials, including transition metal dichalcogenides (TMDs) such as molybdenum disulfide \cite{Mak2010,CLEE}. 

With the demands of modern semiconductor technology, precise nanoscale control of 2D material properties is fervently pursued. The alteration of nanoscale geometry and the introduction of structural defects or strain to these materials can be used to modulate their electronic and optical properties \cite{Ataca2011, Ghorbani-Asl2013, Zhou2010, Zhou2013, Lucchese2010, Cancado2011, Zhao2016}. Ion irradiation has an illustrious record over many decades in the scalable and precise defect engineering of materials \cite{Nastasi1996}. Modern ion irradiation techniques with sub-nanometre probe sizes have demonstrated sub-10 nm precision in fabricating nanoribbons from various materials \cite{Bell2009,Kalhor2014,Fox2013,Fox2015,Naitou2015,Archanjo2014,Nair2012}.

Ultimate modification precision is the convolution of the impact probe and the interaction volume. Given the superlative confinement of 2D materials in the out-of-plane direction, and the confinement of modern ion beams in the in-plane direction, it is possible to restrict the ion-specimen interaction to an exceptionally small volume. Such methods have been used to control doping by implantation and to introduce precise quantities of defects \cite{Fox2015, Nipane2016, Cancado2011, Mignuzzi2015, Pollard2014a}. He$^+$ irradiation of graphene encapsulated in hexagonal boron nitride has been used to introduce n-type doping \cite{Nanda2015} while dose-dependent irradiation of TMDs has introduced pseudo-metallic phases in the monolayer limit \cite{Fox2015,Stanford2017b}.

This paper reports a novel experimental comparison of two ion bombardment species, He$^+$ and Ne$^+$, at high energy. It is also one of very few reports to carry out explicit comparisons between supported and freestanding 2D materials. The introduction of defects by ion irradiation is shown to be highly precise, and Raman spectroscopy proves to be a highly versatile and sensitive characterisation method for these defective monolayer materials.

\subsection{\label{sec:level2}Defect Engineering by Ion Irradiation in 2D Materials}
Defects in 2D materials such as a graphene and MoS$_2$ have been introduced by a range of energetic species. For the application of such ion irradiation methods to defect engineering, the defect yield per ion and the types of defect introduced must be known. The mass and energy of the primary ion species both play a substantial role in determining the average defect yield and the nature of the dominant defect type. With increased mass of energetic noble gas ions, both the defect yield and size in graphene are expected to increase \cite{Lehtinen2010,Lehtinen2011}. For high energy ions the defect yield decreases with increasing energy while the defect size increases.

However, defects induced in supported 2D materials are created not just by the primary ion beam. While a substrate is expected to lower the damage probability per ion for low energy ions, the converse is true for high energy ions as the substrate facilitates backscattering of the primary beam and thus provides a source of energised secondary particles \cite{zhao2012}. \citeauthor{zhao2012} found that irradiation damage was enhanced in supported graphene (compared to freestanding) for energies above 5 keV for Ar$^+$ (M=18) and 3 keV for Si$^+$ (M=14) \cite{zhao2012}. Excited substrate particles have a much lower energy than the primary ions which suggests that they would create defects of greater size.

\citeauthor{Lucchese2010} used low energy (90 eV) Ar$^+$ ions to alter the average distance between defects, $L_D$, in graphene \cite{Lucchese2010}. In these studies $L_D$ was calculated based on the expected density of defects, $\sigma$, which is estimated from the irradiation dose, $S$, in ions per unit area. The approximation that $\sigma \simeq S$ was used which assumes a random distribution of ions, yielding:\cite{Pollard2014a,Mignuzzi2015,Lucchese2010}  
\begin{equation}
L_D = \frac{1}{\sqrt{\sigma}}
\label{eq:ldnotcorrected}
\end{equation}
This is a valid assumption for ions with a suitable cross-section for creating a single carbon vacancy defect. The 90 eV Ar$^+$ ions applied to graphene by \citeauthor{Lucchese2010} are one such example due to their relatively large mass and low energy. The vast majority of those ions do indeed interact with surface carbon atoms but due to the low energy can only remove one carbon atom each.

\subsection{\label{sec:level2}Local Activation Model for Raman Spectra of Defective 2D Materials}
The Raman spectra of graphene have been extensively studied \cite{Ferrari2006a, Ferrari2007, Wang2008, Lucchese2010, Cancado2011, Beams2011, Ferrari2013a, Pollard2014a, Hang2014} with labelled characteristic peaks: $2D$ at $\sim$2640 cm$^{-1}$ and $G$ at $\sim$1583 cm$^{-1}$. In the Raman spectra of defective graphene, the disruption to normal selection rules also allows the detection of two additional peaks: $D$ at $\sim$1322 cm$^{-1}$, and $D'$ at $\sim$1600 cm$^{-1}$ \cite{Ferrari2006a, Malard2009}. Furthermore, increasing structural disorder also causes the $G$ peak to consistently broaden. $L_D$ in graphene has been related to the ratio of the intensity of the $D$ peak ($I_D$) to the intensity of the $G$ peak ($I_G$) and a version of that relationship is given by\citeauthor{Lucchese2010}:

\begin{equation}
\begin{aligned}
& \frac{I_D}{I_G} =\\{} 
& C_A \frac{(L^2_S+2r_SL_S)}{(L^2_S+2r_SL_S-r^2_S)}\left[e^{-\pi r^2_S/L_D^2 }-e^{-\pi (L^2_S+2r_SL_S)/L_D^2}\right] \\
& + C_S\left[1-e^{-\pi r^2_S/L_D^2}\right]\
\end{aligned}
\label{eq:Lucchese}
\end{equation}
where $r_S$ and $r_A$ represent the radii of a structurally disordered area created by an ion and the radii of the outer $D$ band-activated area respectively.
$L_S=r_A-r_S$ is the Raman relaxation length for the resonant Raman scattering. The intensity of the $D$ peak is proportional to the total area of crystalline graphene that is activated by local defects. Thus, as $L_D$ becomes low (i.e. the defect density increases and the material becomes less crystalline) the $D$ band intensity falls due to the overlapping of the disordered areas and the decrease of the total $D$-activated area. 

The dispersive effect of the excitation energy, $E_l$, on the ratio of excitation of the $D$ and $G$ bands, is included in the $C_A$ parameter, being a measure of the maximum possible value of the $I_{D}$/$I_{G}$ ratio \cite{Cancado2007}. Where $E_l$ is stated in eV, $C_A$ has been given experimentally by: \cite{Cancado2011}

\begin{equation}
\begin{aligned}
C_A = (160\pm 48)\times E_l^{-4}
\end{aligned}
\label{eq:CA}
\end{equation}

The $C_S$ parameter is the value of the I$_{D}$/I$_{G}$ ratio in the highly disordered limit and it is important in the large defect density regime, $L_D \le r_S$ \cite{Lucchese2010,Cancado2011}.
Typically, three stages are discussed in the evolution of the relationship described by equation \ref{eq:Lucchese}. The first stage begins with pristine graphene. As isolated defects initially appear in the crystalline lattice a rising $D$ peak is observable, increasing $I_{D}/I_{G}$. The second stage features red-shifting and continued broadening of the $G$ peak and a now diminishing $D$ peak. It is reached when defects coalesce and carbon valence declines \cite{Pollard2014a,Cancado2011,Hang2014,Lucchese2010}.  The third stage is marked by the transition of the specimen to amorphous carbon bearing limited resemblance to the original graphene.

Raman spectroscopy has been employed extensively in the characterisation of MoS$_2$ in various forms \cite{Wieting1971,Chen1974,Windom2011,Livneh2015,  Stacy1985, Frey1999, Rice2013} including monolayer (of polytype 1H, point group D$_{3H}$) which has the labelled characteristic peaks: $E'$ at $\sim$383 cm$^{-1}$ and $A'_1$ at $\sim$401 cm$^{-1}$. In defective material, an additional peak, the $LA(M)$ at $\sim$227 cm$^{-1}$ is found \cite{CLEE, Sanchez2011, Li2012, Li2012a}. The $E'$ peak comes from the intralayer, in-plane motion of Mo and S atoms with respect to each other and the $A'_1$ peak comes from the intralayer, out of plane motion of S atoms \cite{CLEE, Li2012, Livneh2015, Ye2015, Mak2012}.

The $LA(M)$ peak appears in nanoparticle/multi-layer samples but exhibits no intensity in pristine monolayer MoS$_2$ \cite{Frey1999, Stacy1985, Livneh2015}. However, it intensifies quickly with increased defect density \cite{Frey1999, Mignuzzi2015}. 
Since it is defect-activated, \citeauthor{Mignuzzi2015} draw an analogy between the $LA(M)$ peak in monolayer MoS$_2$ and the $D$ peak in graphene as both represent a good measure of disorder when normalized \cite{Mignuzzi2015}. The intensity of the $LA(M)$ peak, $I(LA)$, normalized to that of either the $E'$ peak, $\frac{I(LA)}{I(E')}$, or the $A'_1$ peak, $\frac{I(LA)}{I(A'_1)}$, is related to the inverse square of $L_D$ by:

\begin{equation}
\frac{I(LA)}{I(X)}=\frac{C(X)}{L_D^2}
\label{eq:mos2}
\end{equation}
In the case where the Raman spectrum is acquired with a 532 nm laser, the following constants were reported from the fitting of experimental data: $C(E') = 1.11\pm0.08$ nm$^2$ and $C(A'_1) = 0.59\pm0.03$ nm$^2$. $X = E'$ or $A'_1$ depending on the peak studied. During defect engineering, the increase of these intensity ratios is attributed to two concomitant factors: (i) an increase in the absolute intensity of the defect-activated $LA(M)$ peak, and (ii) a decrease in the intensity of the $E'$ and $A'_1$ peaks attributed to the ablation of the specimen \cite{Mignuzzi2015}.

\section{Experiment}
\subsection{Preparation of Monolayer 2D Materials} 
Chemical vapour deposition (CVD) was used to grow the graphene sample on copper foil \cite{Kumar2011}. The graphene was transferred onto a Si substrate using a polymer-assisted process as outlined previously \cite{Kumar2011, Hallam2014}. The Si substrate had arrays of holes with a diameter of $\sim$2 $\mu$m and depths of $>10$ $\mu$m as pictured in the supplementary information.

MoS$_2$ was also prepared using a CVD technique \cite{OBrien2014}. MoO$_3$ substrates were placed face-up in a ceramic boat with a blank SiO$_2$ substrate face-down on top. This was situated in the centre of the heating zone of a quartz tube furnace, and ramped to 750 $^{\circ}$C under 150 SCCM of Ar flow. Sulfur (S) vapour was then produced by heating S powder to $\sim$120 $^{\circ}$C in an independently controlled upstream heating zone of the furnace, and carried downstream to the MoO$_3$ for a duration of 20  min. After this, the furnace was held at 750~$^{\circ}$C for 20 min, then cooled down to room temperature. Monolayer flakes of MoS$_2$ with a typical triangular shape could then be identified on the SiO$_2$ surface by optical contrast. 

\subsection{Irradiation with 30 keV He$^+$ and Ne$^+$}
The \textit{Zeiss ORION NanoFab} microscope was used to irradiate arrays of $5 \times 5$ $\mu m^2$ regions in graphene and MoS$_2$ with He$^+$ and Ne$^+$ at an energy of 30 keV and an angle of incidence of 0$^{\circ}$. These regions received doses ranging from 1.5$\times$10$^{11}$ to 1$\times$10$^{16}$ Ne$^+$ cm$^{-2}$ or 1$\times$10$^{13}$ to 1$\times$10$^{17}$ He$^+$ cm$^{-2}$. 
The beam was defocused ($\sim$10s of nm) to ensure a uniform distribution of ions and the sample was irradiated at the desired dose. 1 pA beam current and 10 nm pixel spacing were used. The beam dwell time at each pixel and/or the number of repeats at each position were varied to achieve the desired dose. The chamber pressure was of the order 3 $\times$ 10$^{-7}$ Torr. 

\subsection{Raman Spectroscopy}
Raman spectroscopy was carried out on graphene with a \textit{Horiba Jobin-Yvon} system (633 nm laser) with a 1200 lines/mm diffraction grating and a 100$\times$ objective aperture (NA=0.66) (laser spot size was $\sim$0.7 $\mu$m). These spectra were comprised of 10 acquisitions, each of 1 s duration at a single point for each irradiated region.
 
Raman spectroscopy was carried out on MoS$_2$ using a \textit{WITec Alpha 300R} system (532 nm laser) with a 1800 lines/mm diffraction grating and a 100$\times$ objective (NA=0.95) (laser spot size was $\sim$0.3 $\mu$m). Raman maps were generated by taking four spectra per $\mu$m in both x and y directions over large areas \cite{Delhaye1975}. The acquisition time was 0.113 s. The spectra from a desired region were acquired by averaging.

For both materials, the laser power was $\sim$1 mW to minimise sample damage. Peaks in the Raman spectra were fitted with Lorentzian functions for graphene. For MoS$_2$, Gaussian functions were used for the $E'$ and $A'_1$ peaks and Lorentzians for the region around the $LA(M)$ peak (demonstrated in the supplementary information). Error bars, where used and unless otherwise stated, are the largest of either the instrumental ($1$ cm$^{-1}$) or the fitting error as acquired from the \textit{fityk} software package, which uses a weighted sum of squared residuals to measure agreement between the fit and the data \cite{Wojdyr2010}.

\section{Results \& Discussion}
\subsection{Graphene}

\begin{figure*}
\centering
\subfloat[He$^+$ Irradiated Freestanding Graphene]{\label{ASpectraGrapheneHeF}\includegraphics[width=3.4in]{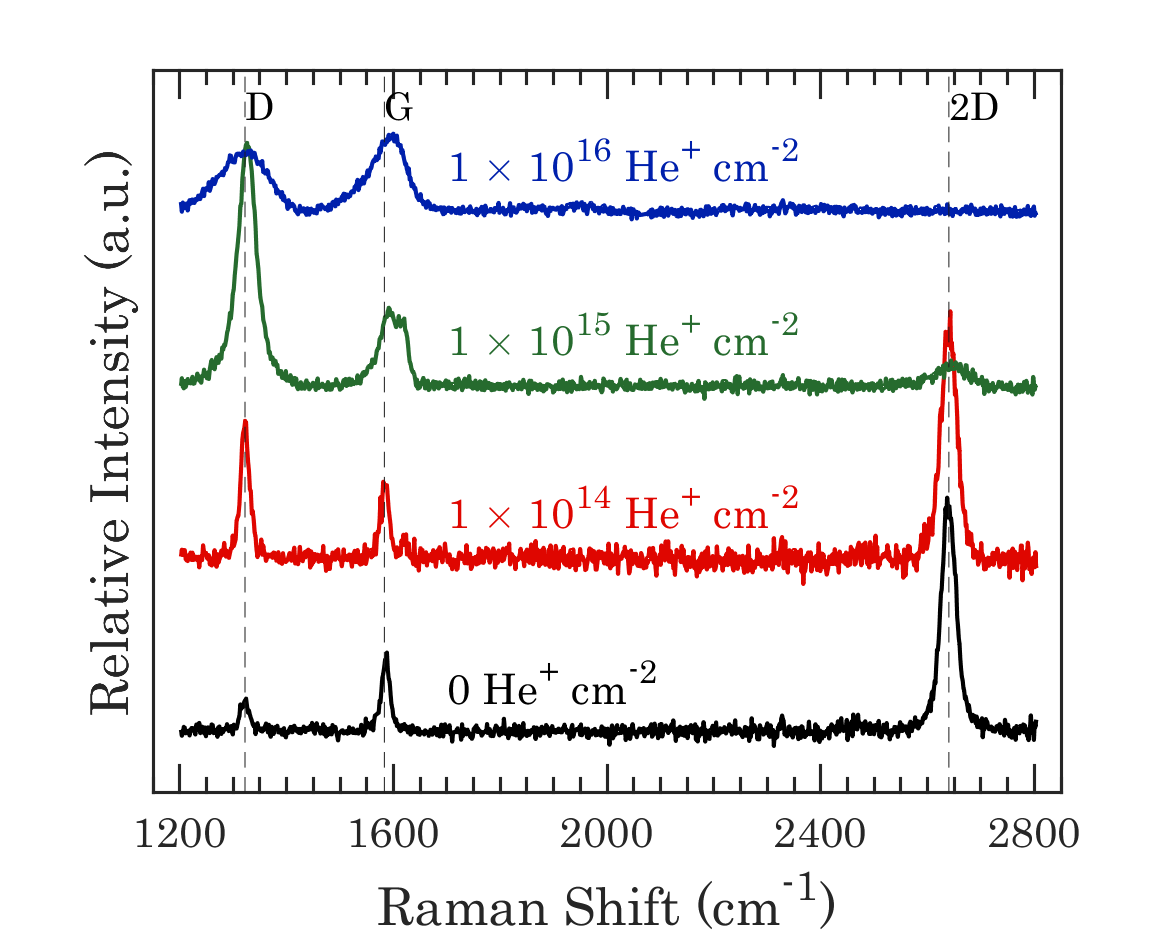}}
\subfloat[He$^+$ Irradiated Supported Graphene]{\label{ASpectraGrapheneHeS}\includegraphics[width=3.4in]{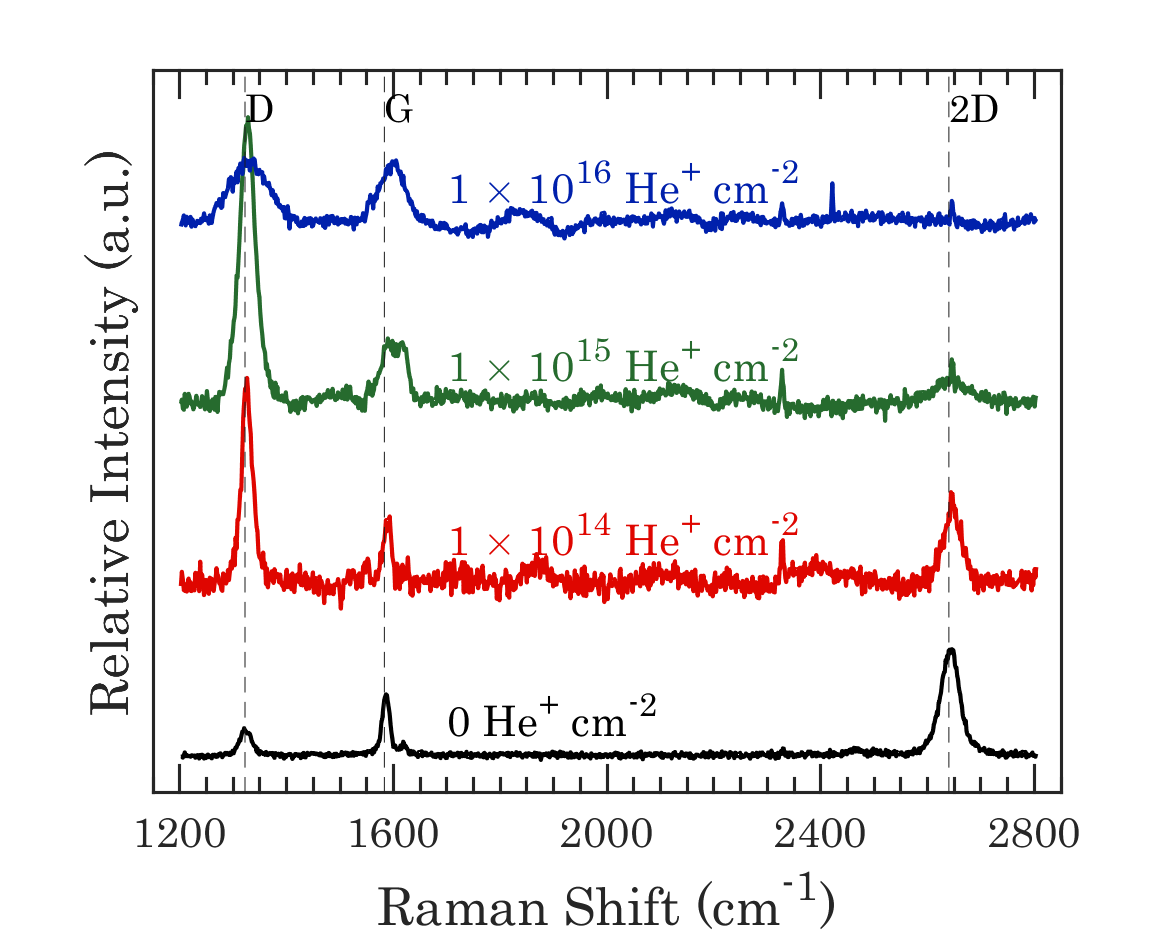}}\qquad
\subfloat[Ne$^+$ Irradiated Freestanding Graphene]{\label{ASpectraGrapheneNeF}\includegraphics[width=3.4in]{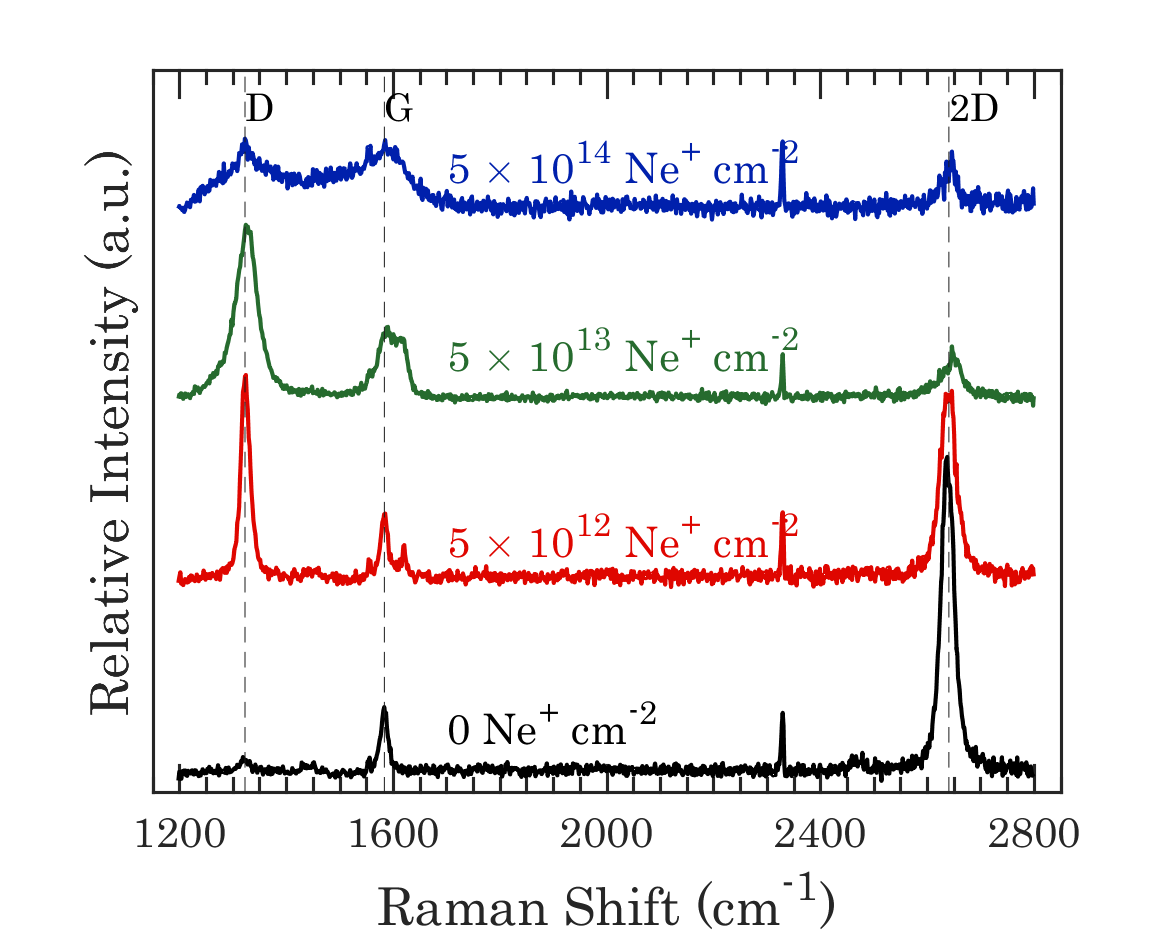}}
\subfloat[Ne$^+$ Irradiated Supported Graphene]{\label{ASpectraGrapheneNeS}\includegraphics[width=3.4in]{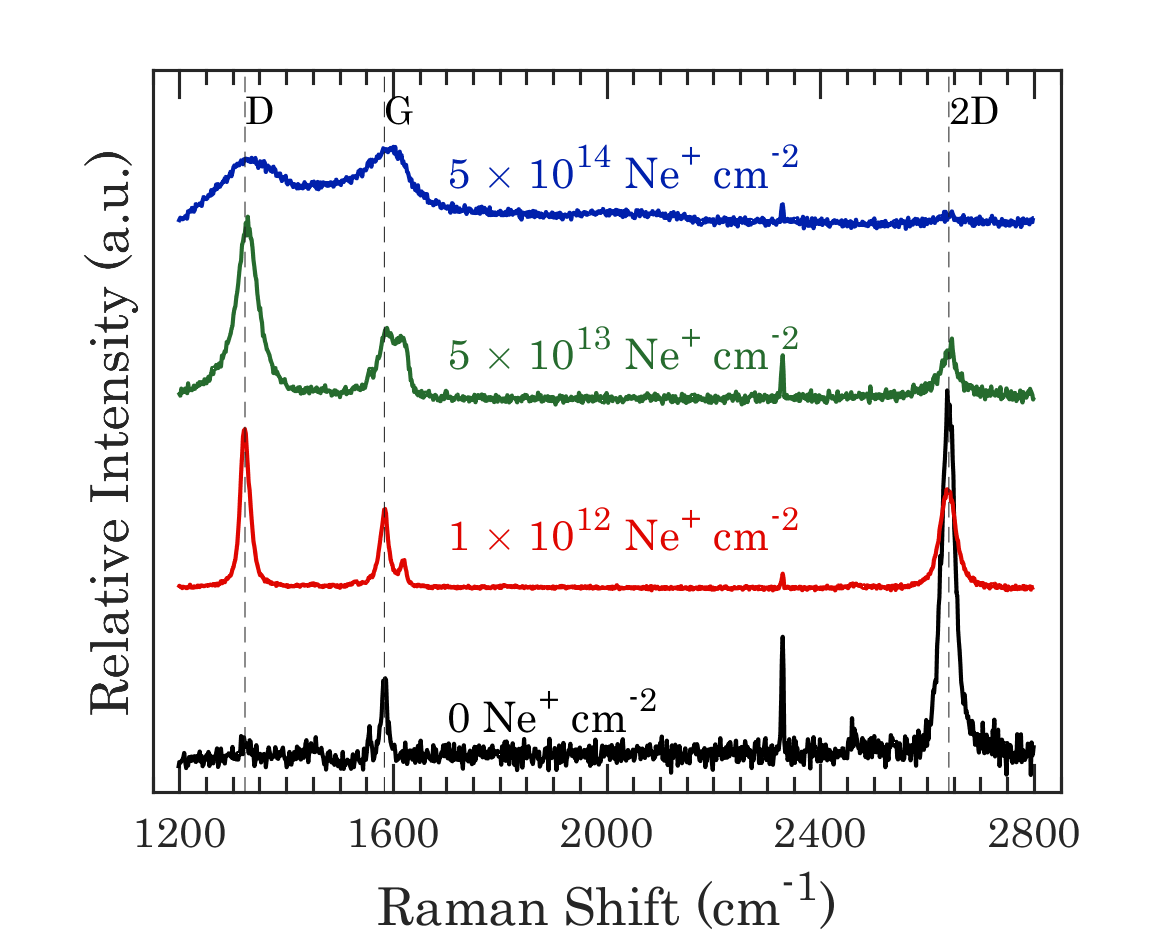}}
\caption{Representative selection of graphene Raman spectra excited by a 633 nm laser and irradiated with ions at 30 keV with a 0$^{\circ}$ angle of incidence. (a) freestanding graphene irradiated with He$^+$, (b) supported graphene irradiated with He$^+$, (c) freestanding graphene irradiated with Ne$^+$, (d) supported graphene irradiated with Ne$^+$. The evolution of the spectra with increased ion dose is shown ascending from the bottom in black to the top in blue. The spectra are normalized to the maximum of the $G$ peak.} 
\label{fig:GrapheneSpectra}
\end{figure*}

In Fig. \ref{fig:GrapheneSpectra}, four sets of Raman spectra are presented, labelled as follows: (a) He$^+$ irradiated freestanding graphene, (b) He$^+$ irradiated supported graphene, (c) Ne$^+$ irradiated freestanding graphene and (d) Ne$^+$ irradiated supported graphene. The spectra obtained from the non-irradiated regions of both supported and freestanding samples are shown in black and agree with the literature for monolayer graphene \cite{Ferrari2013a,Ferrari2006a}.

In each subplot, multiple spectra are shown with dose increasing in ascent from the bottom. The first spectrum (black) represents the non-irradiated graphene. In the second spectrum (red), the defect-activated $D$ peak (at $\sim$1322 cm$^{-1}$) can be observed to have increased in intensity relative to the $G$ band intensity. The third spectrum (green) shows a very intense $D$ peak. Finally, in the fourth spectrum (blue) the material is amorphous carbon with little to no remaining crystallinity. 

\begin{figure*}
\centering
\subfloat[Evolution of Width of the $G$ Peak with Ion Dose]{\label{FWHM-Both}\includegraphics[width=3.40in]{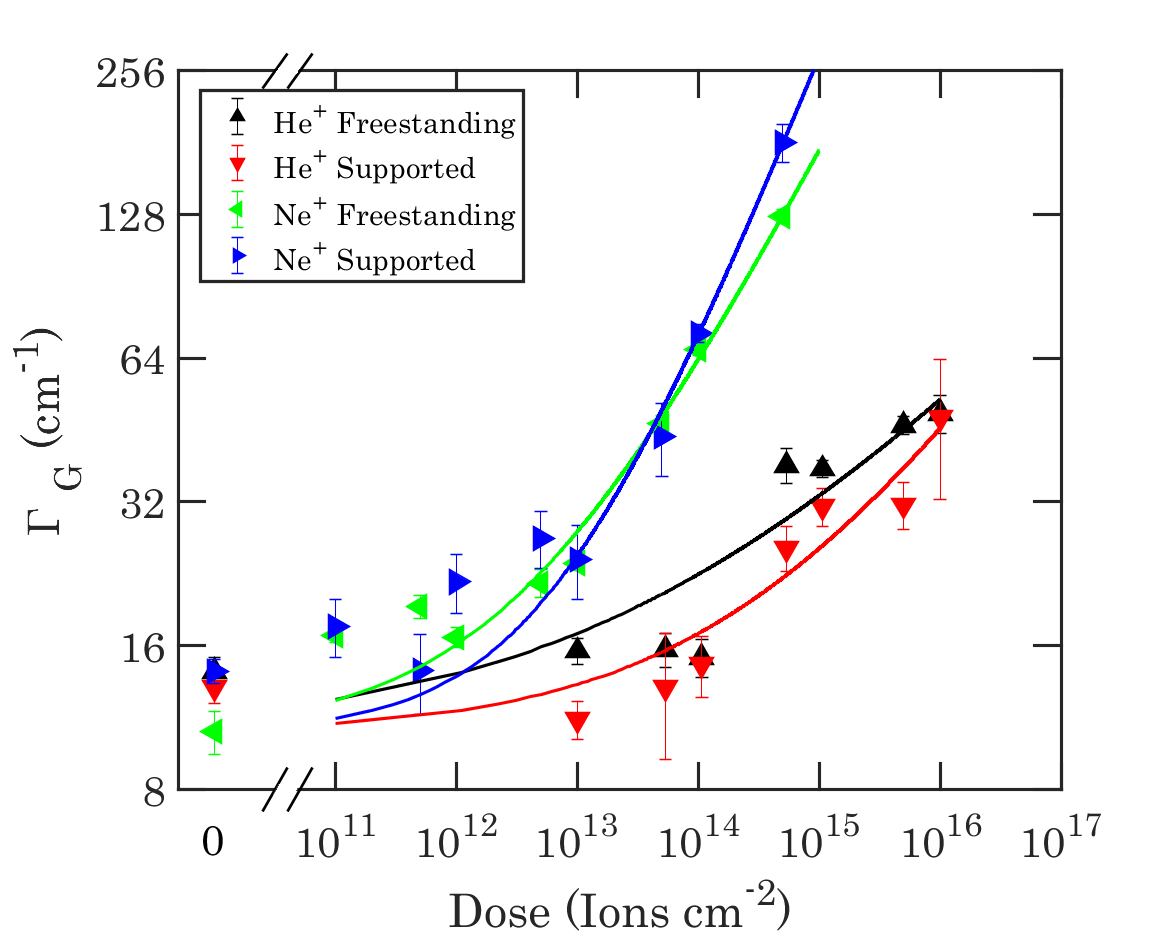}}\hspace{0.5cm}
\subfloat[Evolution of I$_{D}$/I$_{G}$ Ratio with Ion Dose]{\label{AGrapheneIDIGdose}\includegraphics[width=3.40in]{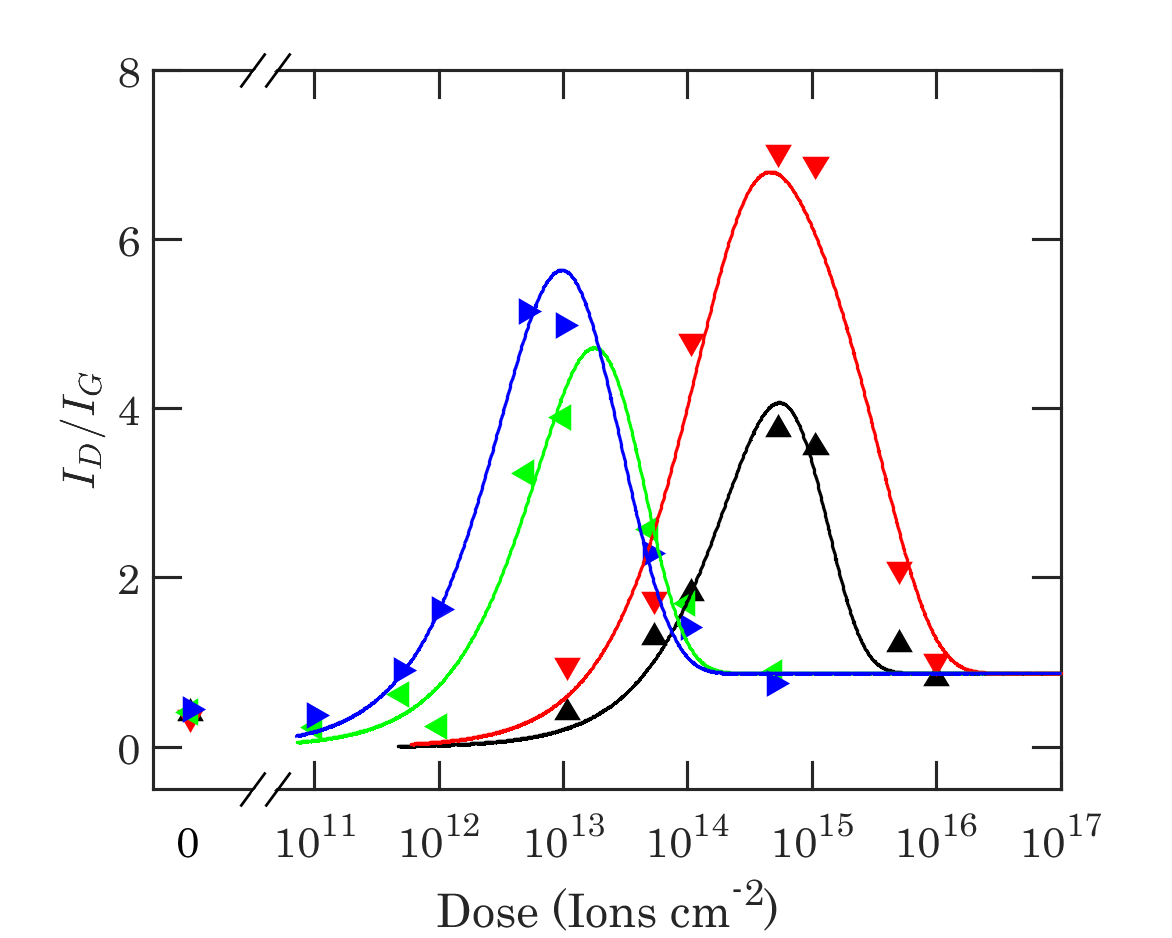}}
\caption{(a) Evolution of full width at half maximum of Lorentzian fits to the $G$ peak, $\Gamma _G$, as a function of dose for He$^+$ and Ne$^+$ for both freestanding and supported graphene. (b) The evolution of the I$_{D}$/I$_{G}$ ratio of graphene with irradiation dose. A version of equation \ref{eq:Lucchese} modified for dose has been fitted to each of the four data sets. The legend in (a) applies to both graphs.} 
\label{fig:FwhmIdigGraphene}
\end{figure*}

The width of the $G$ peak ($\Gamma_G$) increases as a function of ion dose as shown in Fig. \ref{fig:FwhmIdigGraphene}(a) for both He$^+$ and Ne$^+$. For each type of sample and choice of ion this represents an unequivocal increase in structural disorder. While the trends for both ions are similar, the incidence of the Ne$^+$ species causes changes in the $G$ peak to occur at ion doses which are lower by between one and two orders of magnitude than those of He$^+$.
This is due to the increased mass of Ne$^+$ ions and an enhanced milling capability at the incident surface compared to He$^+$ \cite{Rahman2012}. It should be noted that the difference between the effects of ion irradiation on $\Gamma_G$ for supported and freestanding graphene is very small here and difficult to separate from the direct effect of the substrate/suspension on the $G$ peak \cite{Berciaud2008,Ferrari2013a}. The fit to the data is of the form:

\begin{equation}
\Gamma _G = \Gamma _{G0} +bS^c
\label{eq:width}
\end{equation}

Where $\Gamma _{G0}$ is the width of the G peak in pristine graphene, $b$ is a fitting parameter related to the defect yield, $S$ is the ion dose and $c$ is a fitting parameter related to the effect of defect density on the $G$ peak. $c$ is found to be less than 1 in all cases, suggesting that the relationship between dose and $\Gamma _G$ is sub-linear. 
In Fig. \ref{fig:FwhmIdigGraphene}(b), the evolution of the $I_{D}/I_{G}$ ratio against dose is displayed for both He$^+$ and Ne$^+$ and for both freestanding and supported graphene. Progression through the three previously discussed stages is observed. $I_{D}/I_{G}$ of the supported graphene is noted to rise faster and reach a much higher maximum than the freestanding material. The maximum $I_{D}/I_{G}$ ratio is also observed to be lower for freestanding Ne$^+$ irradiated graphene than He$^+$ irradiated graphene but larger for supported Ne$^+$ irradiated graphene than freestanding Ne$^+$ irradiated graphene. This is discussed in relation to defect sizes later.

\begin{table}
    \centering
    \begin{tabular}{ c c c c c c c}
     Ion & Graphene & $\gamma _{C(SRIM)}$ &$\gamma _{C(MD)}$& $\alpha _{C(fit)}$& $\alpha _{(MD)}$
     \\ \hline
     He$^+$ & Freestanding & $0.010$ & $0.006$ & $0.016$ & 0.049
     \\ 
     He$^+$ & Supported    & $0.137$ & --      & $0.024$ & --
     \\
     Ne$^+$ & Freestanding & $0.236$ & $0.156$ & $0.414$ & 1.117
     \\ 
     Ne$^+$ & Supported    & $3.15$  & --      & $0.965$ & --

     \\ 
     &
    \end{tabular}
    \caption{Sputtering yields ($\gamma$) and defect per incident ion ($\alpha$) values of carbon from graphene irradiated with He$^+$ and Ne$^+$ at 30 keV and 0$^{\circ}$ angle of incidence for the four different arrangements discussed in the main text.
    $\gamma _{C(SRIM)}$ is the sputtering yield calculated using SRIM, 
    $\gamma _{C(MD)}$ is the sputtering yield calculated using the online simulation from \citeauthor{Lehtinen2010} \protect\cite{Lehtinen2010, Lehtinen2011},
    $\alpha _{C(fit)}$ is the defect per ion value calculated from the fits shown in Fig. \ref{fig:GrapheneIDIG} and
    $\alpha _{C(MD)}$ is the defect per ion value calculated using the online simulation.}
    \vspace{0ex}
    \label{table:yields}
\end{table}

\subsubsection{Defect Probabilities}
For lighter and/or higher energy ions such as those used in this work, the assumption made by \citeauthor{Lucchese2010} that one ion produces one defect may no longer be valid. In those cases, the average distance between defects becomes:
\begin{equation}
L_D = \frac{1}{\sqrt{\alpha S}}
\label{eq:ld}
\end{equation}
Where $\alpha$ is the defect per ion yield and is distinguished from the sputtering yield, $\gamma$, because not all defects causing local activation of Raman modes need necessarily be a vacancy.
The sputtering yields of graphene were calculated using stopping and range of ions in matter (SRIM) simulations ($\gamma _{C(SRIM)}$) as detailed in the supplementary information \cite{ziegler2010srim,Ziegler}. 
In calculating the yield of carbon atoms from graphene, SRIM accounts for those carbon atoms removed by primary ions, backscattered ions and secondary particles. 
The four scenarios applied in experiment were simulated: (i) freestanding, irradiated with He$^+$, (ii) supported, irradiated with He$^+$, (iii) freestanding, irradiated with Ne$^+$, and (iv) supported, irradiated with Ne$^+$. These values are presented in table \ref{table:yields} alongside values calculated using the molecular dynamics derived simulation package of \citeauthor{Lehtinen2010} \cite{Lehtinen2010}. The sputtering yields calculated using SRIM for each ion are about 10 times larger in supported than freestanding graphene. This indicates that secondary particles from the substrate have a significant impact on the rate of defect introduction to the graphene layer.  

\begin{figure*}
\centering
\subfloat[Freestanding and Supported Graphene Irradiated with He$^+$]{\label{GrapheneHe}\includegraphics[width=3.40in]{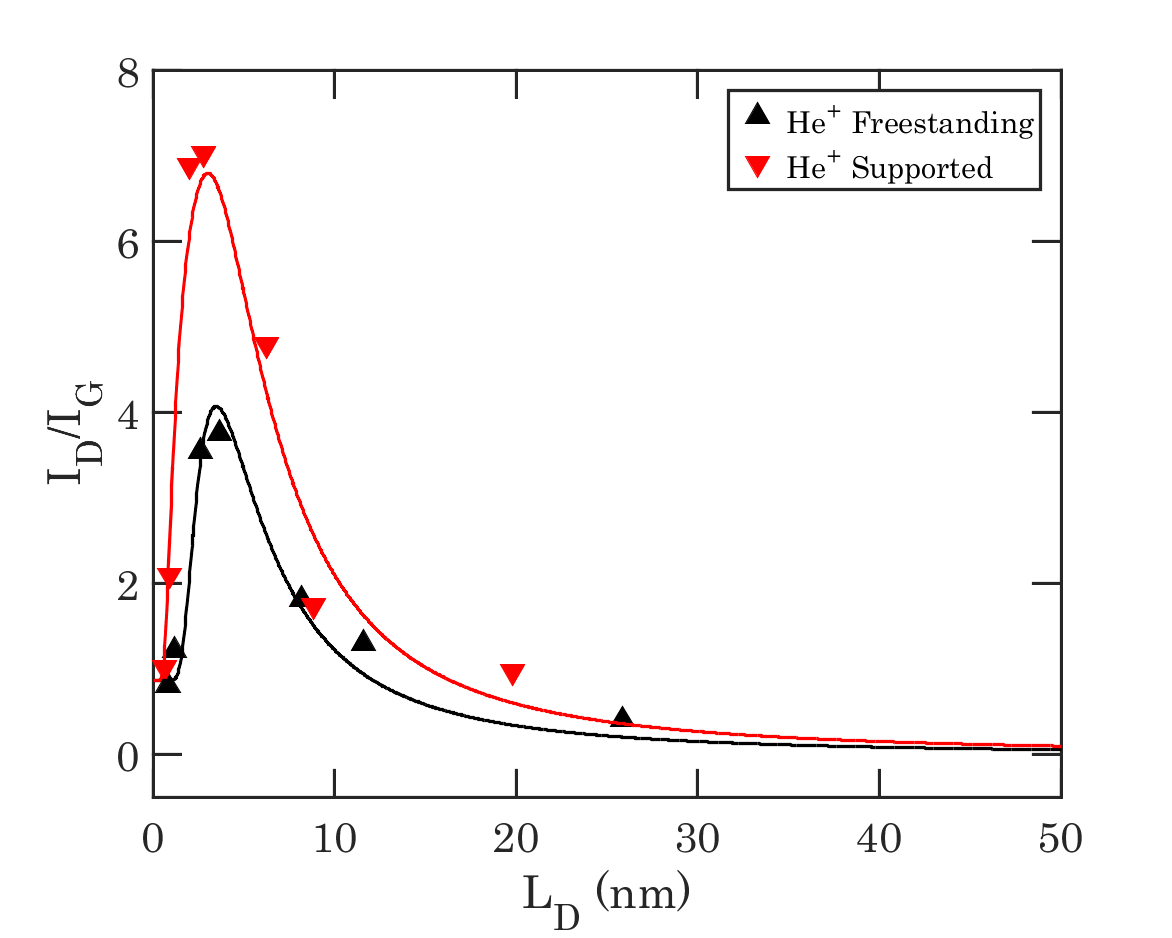}}\hspace{0.5cm}
\subfloat[Freestanding and Supported Graphene Irradiated with Ne$^+$]{\label{GrapheneNe}\includegraphics[width=3.40in]{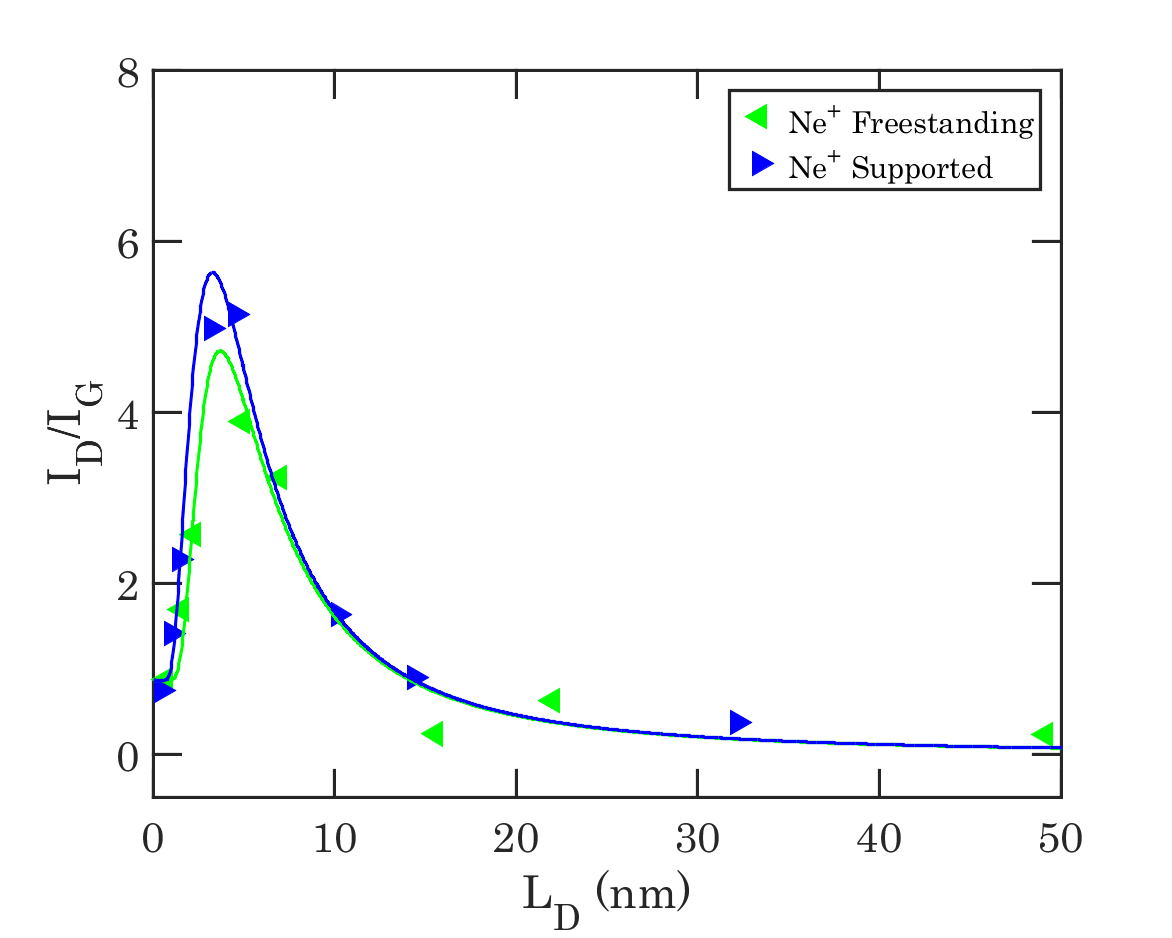}}
\caption{The evolution of the I$_{D}$/I$_{G}$ ratio of freestanding and supported graphene with $L_D$ for (a) He$^+$ and (b) Ne$^+$. The points represent experimental data and the lines are from a fit to equation \ref{eq:Lucchese} \protect\cite{Lucchese2010}.}
\label{fig:GrapheneIDIG}
\end{figure*}

The process of fitting equation \ref{eq:Lucchese} to our data involved using initial values from previous studies. The process is described in detail in the supplementary information.
Equation \ref{eq:Lucchese} was fitted to the four data sets in Fig. \ref{fig:GrapheneIDIG}(a)(b), showing the close agreement of the experimental data and the fitted equation. 
The values for $\alpha _C$ obtained by fitting, $\alpha_{C(fit)}$, are shown in table \ref{table:yields} alongside their corresponding seed values, the sputtering yield found using SRIM, $\gamma_{C(SRIM)}$. For comparative purposes, two further values are included which are applicable to freestanding graphene only. $\gamma_{C(MD)}$ is the sputtering yield and $\alpha _{C(MD)}$ is the defect per ion value calculated using the molecular dynamics-derived online simulation of \citeauthor{Lehtinen2011} \cite{Lehtinen2010, Lehtinen2011}.
The sputtering yield of He$^+$ on freestanding graphene has also been measured experimentally before using a single pixel exposure to completely mill through a graphene layer. The value reported by \citeauthor{Buchheim2016} is $\gamma _C = 0.007$, which is in close agreement with atomistic simulations \cite{Buchheim2016}. Given that this omits defects that do not involve sputtering, it is expectedly smaller than the value of $\alpha _{C(fit)}$ obtained in this work. Similarly, in the case of Ne$^+$ on freestanding graphene, we obtain a value of $\alpha _{C(fit)}$ which is more than twice the value of $\gamma _{C(MD)}$. 
The obtained result, that $\alpha _C$ is consistently larger for supported rather than freestanding graphene for these irradiation conditions, is thus in keeping with expectations discussed in the introduction. A point to note is that for both He$^+$ and Ne$^+$ in freestanding graphene the experimentally derived $\alpha _{C(fit)}$ is smaller than the computationally derived $\alpha _{C(MD)}$, a discrepancy which is worthy of future study beyond this work. 

\begin{figure*}
\centering
\subfloat[He$^+$ Irradiated MoS$_2$]{\label{ASpectraMoS2HeEALa}\includegraphics[width=3.40in]{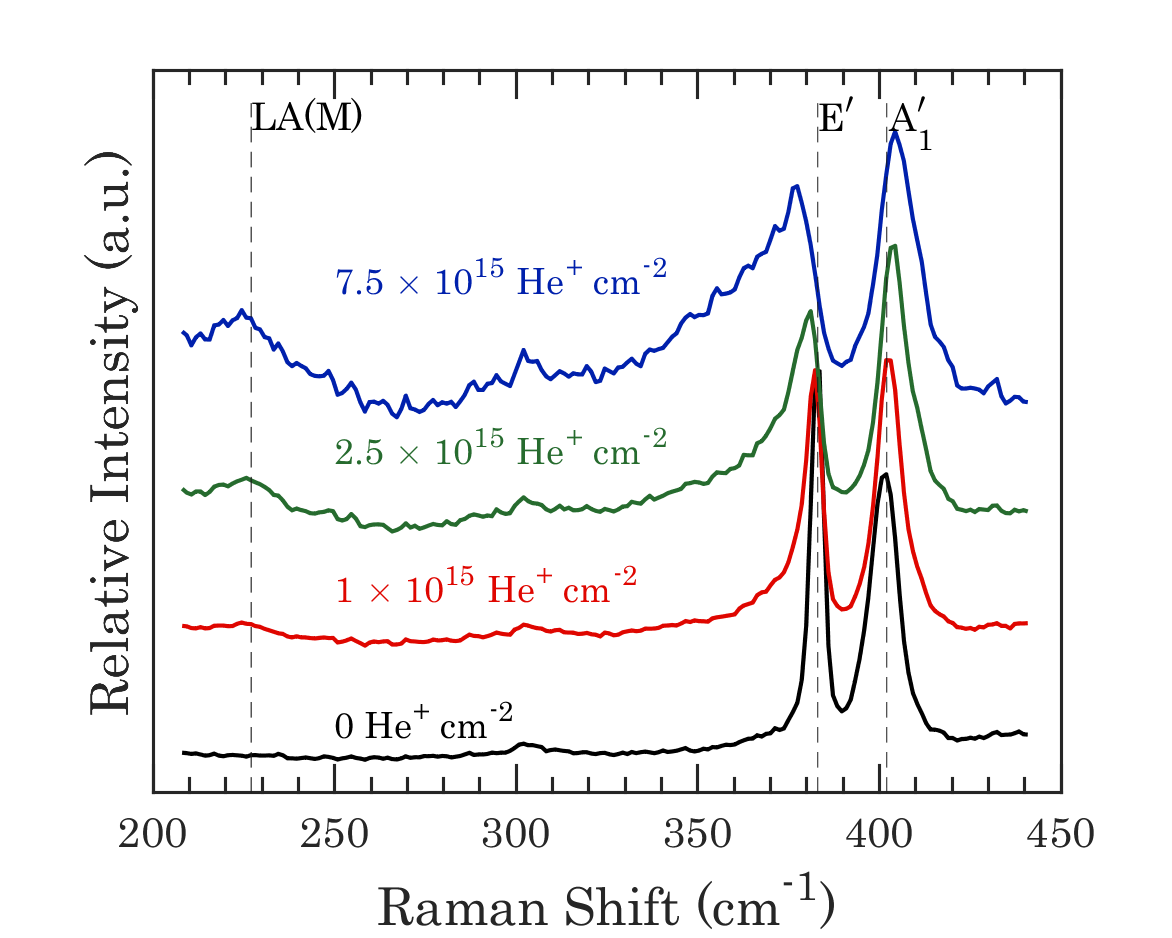}}\hspace{0.5cm}
\subfloat[Ne$^+$ Irradiated MoS$_2$]{\label{ASpectraMoS2NeEALa}\includegraphics[width=3.40in]{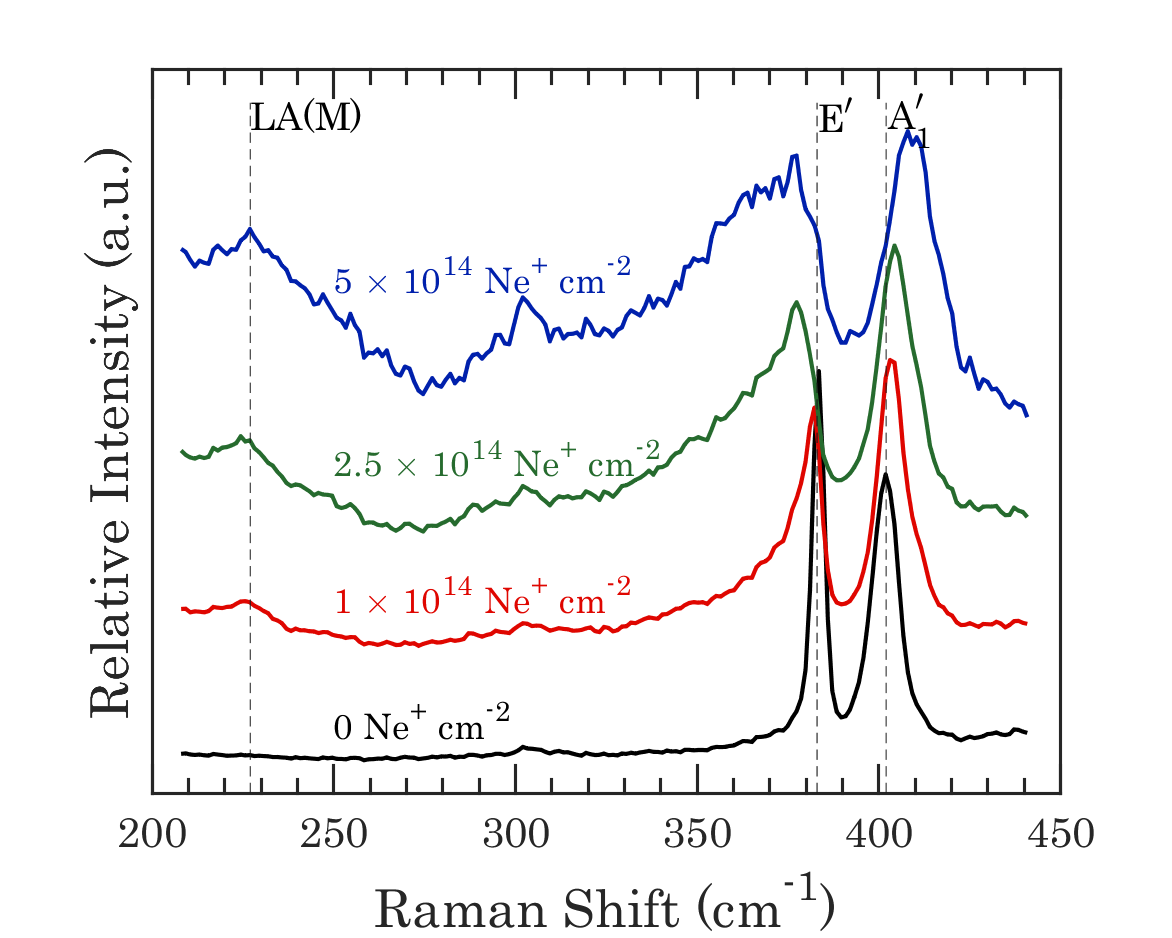}}
\caption{A representative selection of Raman spectra of MoS$_2$ with increased ion doses ascending from pristine at the bottom (in black) to the highest dose at the top (in blue). A 532 nm laser was used as the excitation source. Material irradiated with He$^+$ is shown in (a) and material irradiated with Ne$^+$ is shown in (b). Both ion species had an incident energy of 30 keV. The plots show the $LA(M)$ mode at $\sim$227 cm$^{-1}$ as well as the $E'$ and $A'_{1}$ peaks. The spectra were individually normalized to the $A'_{1}$ peak.} 
\label{fig:figuremos2spec}
\end{figure*}

\subsubsection{Defect Sizes}
\begin{table}
    \centering
    \begin{tabular}{ c c c c c c c c c c c}
     Ion & Graphene &$r_S$ &$r_A$ &$L_{D}\left ( \frac{I_D}{I_G} \right )^{max}$  &$\left ( \frac{I_D}{I_G} \right )^{max}$\\ \hline
     He$^+$ & Free.& ~$1.77$~nm	& ~$2.93$~nm & ~$3.48$~nm& $4.07$\\
     He$^+$ & Supp.& ~$0.62$~nm	& ~$3.20$~nm & ~$3.02$~nm& $6.79$\\
     Ne$^+$ & Free.& ~$1.62$~nm	& ~$3.19$~nm & ~$3.70$~nm& $4.72$\\
     Ne$^+$ & Supp.& ~$1.12$~nm	& ~$3.02$~nm & ~$3.28$~nm& $5.64$\\
     &
    \end{tabular}
    \caption{Key parameters calculated from fitting equation \ref{eq:Lucchese} to graphene experimental data as in Fig. \ref{fig:GrapheneIDIG}. These values are the average defect size ($r_S$), the average radius of $D$ band activated regions ($r_A$), the average distance between defects at which the maxiumum $\frac{I_D}{I_G}$ value occurred ($L_{D}\left ( \frac{I_D}{I_G} \right )^{max}$), and the maximum $\frac{I_D}{I_G}$ value ($\left ( \frac{I_D}{I_G} \right )^{max}$).}
    \vspace{0ex}
    \label{table:params}
\end{table}

Other results of the fitting process are summarised in table \ref{table:params}.
It is noted that the maximum $I_{D}/I_{G}$ value is considerably higher for supported graphene than for freestanding graphene. It is observed from equation \ref{eq:Lucchese} and displayed in Fig. S1 that there is a close relationship between $r_S$ and the maximum of the $I_{D}/I_{G}$ ratio. Depending on many factors, but principally the irradiation species, energy and angle, we can generally expect $r_S$ values between approximately 0.8 and 2.5 nm \cite{Pollard2014a,Gawlik2017}. 
It has been established that larger incident species usually produce larger defects \cite{Pollard2014a,Lehtinen2010,Lehtinen2011}. However, in the freestanding case, $r_S$ is unexpectedly not found to be larger for Ne$^+$ than He$^+$, despite the very different ion mass. This seemingly anomalous He$^+$ behaviour may be related to a similarly unexpected experimental finding by \citeauthor{Gawlik2017} \cite{Gawlik2017}.
Also observable are the substantially higher values for $r_S$ for freestanding than supported graphene. \citeauthor{Lehtinen2010} found that the size of defects introduced by energetic particles typically increases with energy \cite{Lehtinen2010}. While the high energy interaction of the direct beam is expected to cause large defects, it also excites substrate particles which have lower energies than the incident ion. These lower energy atoms can produce numerous, smaller defect sites in the graphene at the surface. 

A comparison of ion species is provided in table \ref{table:mass}. For supported graphene irradiated with 30 keV He$^+$, $r_S$ has been found to be $\sim$0.8-1 nm \cite{Archanjo2014,Hang2014}. The value obtained in this work for He$^+$ is slightly smaller than previous reports \cite{Archanjo2014,Hang2014}. Since the mass of Ne$^+$ is between those of He$^+$ and Ga$^+$, by a simple argument, it might be expected that the corresponding $r_S$ value would be similarly intermediate. Although the result for He$^+$ irradiated freestanding graphene seems to be anomalous as previously discussed, this is indeed true for our supported graphene. Ga$^+$, also incident on supported graphene and also at 30 keV, has been reported to create defects of $r_S=1.6$ nm \cite{Archanjo2014}. Thus the trend for supported graphene is clearly one of increasing defect size with increasing ion mass, $m_a$.

\begin{table}
\centering
    \begin{tabular}{c c c c c}
    &Ion & $r_S$ &$m_a$\\ \hline
        & He$^+$& 0.62 nm & 4\\
	    & Ne$^+$& 1.12 nm & 20.1\\
	    & Ga$^+$& 1.6 nm  & 69.7 & \cite{Archanjo2014}\\
    \end{tabular}
    \caption{
    Comparison of defect size ($r_S$) in supported graphene caused by three ion species of different masses ($m_a$) at 30 keV.
    }
    \vspace{0ex}
    \label{table:mass}
\end{table}

We propose that the larger ion transfers energy more efficiently to substrate atoms (the masses of Si and Ne are very close), and it is these particles, being more efficiently energised than their He$^+$-induced counterparts, which create larger defects in the graphene layer. The variety of $r_S$ values obtained in this work suggests a variety of defect types with different weighting in the four experimental scenarios. Such defects may include single vacancies, double vacancies, complex defects or amorphisation \cite{Lehtinen2010,Lehtinen2011}. This underscores the importance of choosing ion and substrate carefully for both nanofabrication and defect engineering. 


\begin{figure*}
\centering
\subfloat[FWHM of $E'$ and $A'_1$ peaks]{\label{AMoS2FWHM}\includegraphics[width=3.40in]{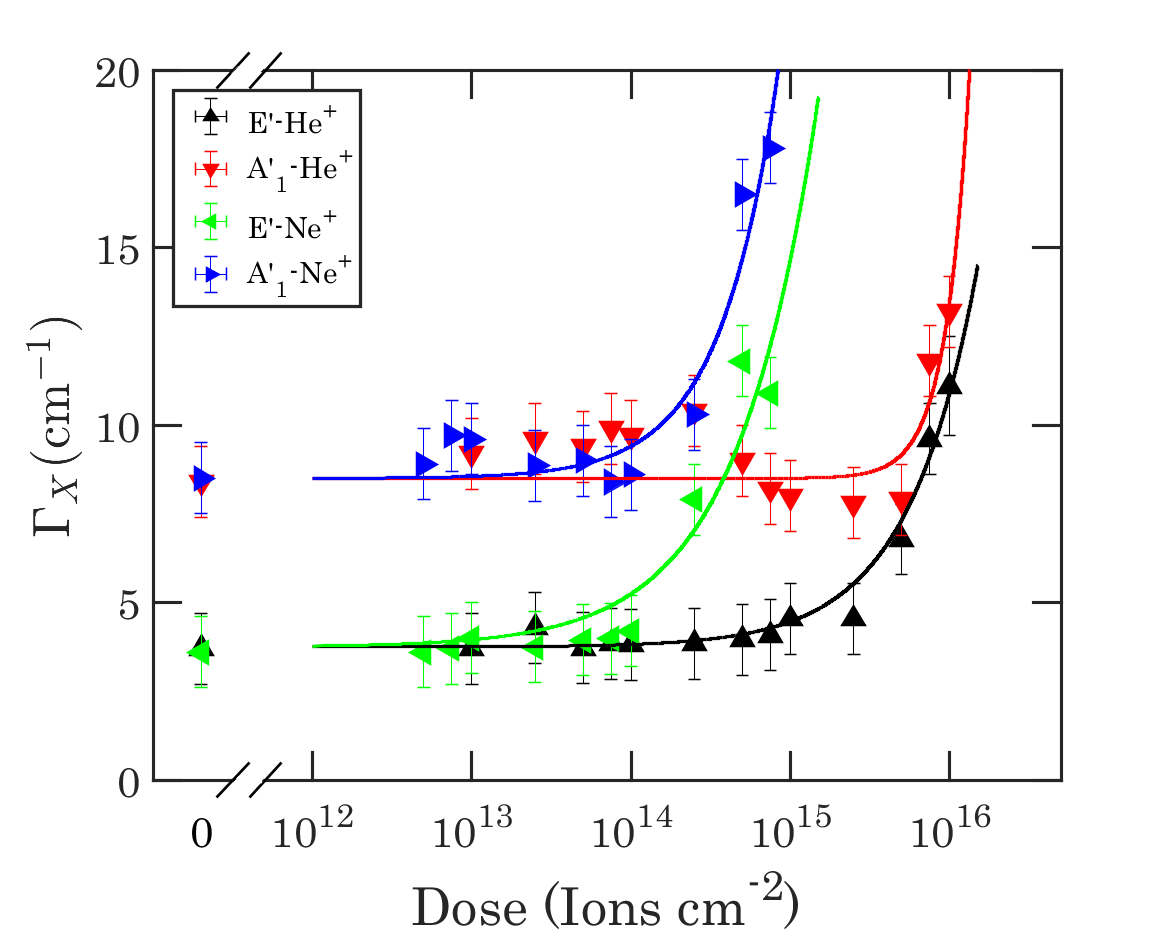}}\hspace{0.5cm}
\subfloat[Positions of $E'$ and $A'_1$ peaks]{\label{AMoS2Centres}\includegraphics[width=3.40in]{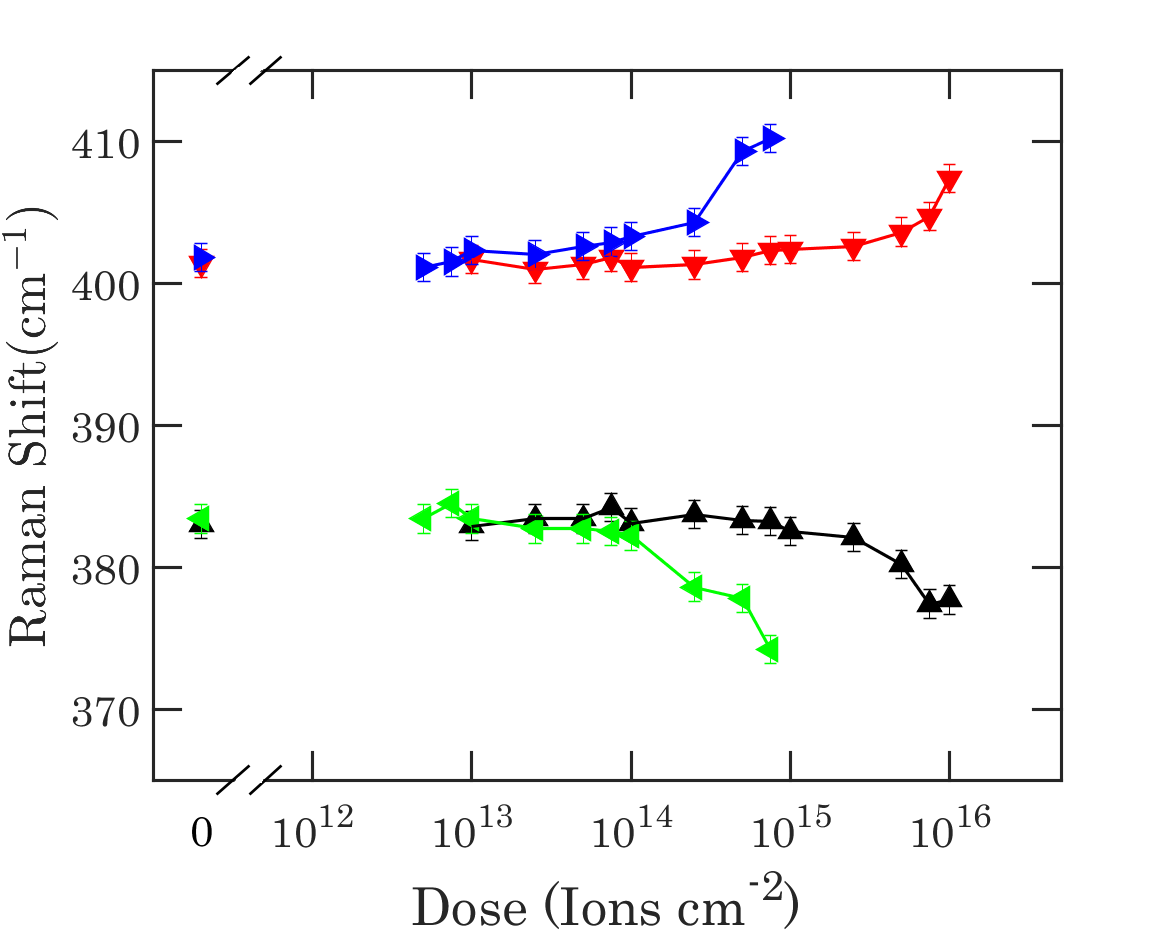}}
\caption{(a) The evolution of the FWHM ($\Gamma_X$) of MoS$_2$ Raman modes as a function of ion dose. The data was fitted with a modified version of equation \ref{eq:width}. (b) The evolution of peak position with ion dose. The legend in (a) applies to both graphs.}
\label{fig:AMoS2widthandcentres} 
\end{figure*}

In Fig. \ref{fig:figuremos2spec} (a)(b), the characteristic $E'$ and $A'_1$ peaks of monolayer MoS$_2$ are marked by dark grey dashed lines.
In the pristine spectra (in black) they are in good agreement with the literature and the small separation of these two peaks ($\sim$18 cm$^{-1}$) is indicative of monolayer MoS$_2$.
With increasing ion dose, quenching and broadening of these two characteristic peaks are observed, reflecting the growing disorder which the ion beams create in the material. The emergence of the $LA(M)$ peak at $\sim$227 cm$^{-1}$(dashed grey line), particularly at high doses of He$^+$ and Ne$^+$, is evident in Fig. \ref{fig:figuremos2spec}. 

The evolutions of width and position of the $E'$ and $A'_1$ peaks are shown in Fig. \ref{fig:AMoS2widthandcentres} (a) and (b) respectively. 
Both peaks are observed to broaden with increasing disorder and the peak positions shift as expected from previous reports \cite{Klein2017}. Broadening begins at a substantially lower dose for Ne$^+$ than for He$^+$ as expected. The $E'$ peak red-shifts and this downward shift in energy is attributed to the introduction of defects causing lattice distortion, similarly to tensile strain \cite{CLEE, Gomez2012, Rice2013, McCreary2016, Li2012, Zhang2013, OBrien2017}. The $A'_1$ peak also blue-shifts for some of the higher doses used, as previously reported \cite{Mignuzzi2015}. 

\begin{figure*}
\centering
\subfloat[]{\label{DoseIntRatios}\includegraphics[width=3.40in]{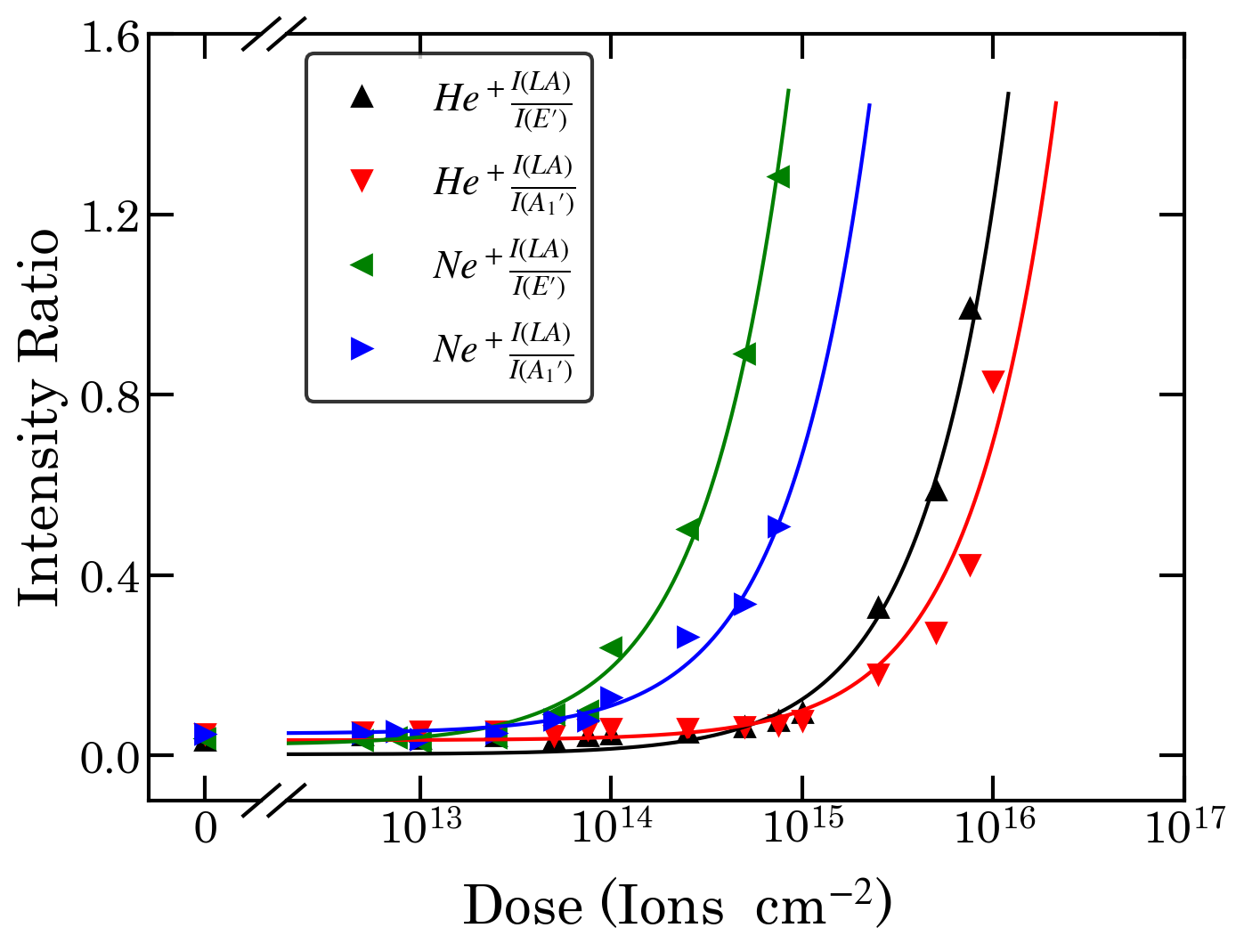}}\hspace{0.5cm}
\subfloat[]{\label{LDIntRatios}\includegraphics[height=2.6in]{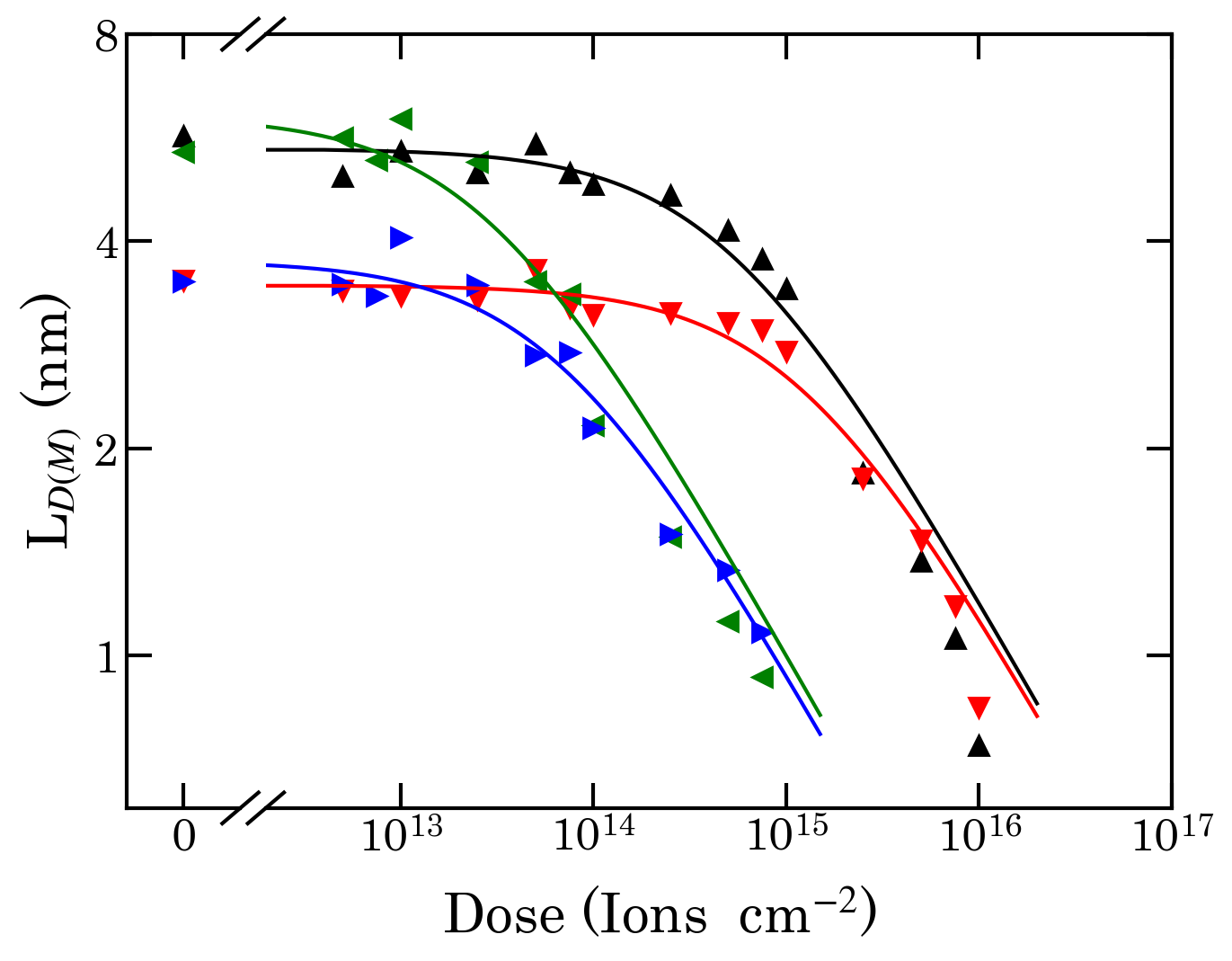}}
\caption{(a) shows the evolution with ion dose of the ratios of the intensity of the $LA(M)$ peak to the intensities of the $E'$ and $A'_{1}$ modes for both He$^+$ and Ne$^+$ irradiated MoS$_2$. (b) shows $L_{D(M)}$ as calculated from the intensity ratios in (a) and using the work of \citeauthor{Mignuzzi2015} \cite{Mignuzzi2015}. The legend in (a) applies to both graphs.}
\label{fig:AMoS2Ratios} 
\end{figure*}

Fig. \ref{fig:AMoS2Ratios} (a) shows the evolution of the intensity ratios extracted from the spectra as a function of dose. With increasing disorder introduced by both ion beams, a sharp increase in the intensity of the $LA(M)$ peak normalized to both the $E'$ and $A'_1$ peaks is observed. Once again, the increased defect yield of Ne$^+$ compared to He$^+$ is highlighted.

\citeauthor{Mignuzzi2015} used equation \ref{eq:mos2} to relate these intensity ratios directly to the average interdefect distance \cite{Mignuzzi2015}. We highlight some caveats to this approach in addition to those present for graphene. Since MoS$_2$ is non-monoatomic, changes in stoichiometry may cause more complex defect-dependent behaviour. Also, \citeauthor{Mignuzzi2015} made an implicit assumption that each ion causes one defect which we do not consider to be a safe assumption for either the $25$ keV Mn$^{+}$ used in their work or indeed the lighter 30 keV ions used in this work \cite{Lucchese2010}. Nonetheless, this form of the defect-activation model and the constants provided previously were applied to our data to calculate the average displacement between defects, $L_{D(M)}$, for both peaks and for both He$^+$ and Ne$^+$ irradiation. In order to calculate the defect yield per ion, $\alpha_M$, the initial defect level is accounted for in an adjusted version of equation \ref{eq:ld}: 

\begin{equation}
L_{D(M)} = \frac{1}{\sqrt{\alpha_M S + \sigma_i}}
\label{eq:ldmos2}
\end{equation}
Where $\sigma_i$ is the defect density in non-irradiated MoS$_2$. These $L_{D(M)}$ values are presented as a function of ion dose in Fig. \ref{fig:AMoS2Ratios} (b). Even the non-irradiated MoS$_2$ has a somewhat low $L_{D(M)}$, not unusual for CVD-grown material.  It is notable that in the high $L_{D(M)}$ range, where the ion dose is small, there is a discrepancy between the values given by the two peaks. Using $\frac{I(E')}{I(LA(M)}$ to calculate $L_{D(M)}$ yields a consistently higher value than using $\frac{I(A'_1)}{I(LA(M)}$. However, the two values approach each other as $L_{D(M)}$ decreases which suggests that the nature of the initial defects may be different to those introduced by ion irradiation. Equation \ref{eq:ldmos2} allows for the extraction of the $\sigma_i$ and $\alpha_M$ values for both peaks and both ions as presented in table \ref{table:mos2results}. The values for $\sigma_i$ are in good agreement between the two ions, though as mentioned already do not agree between the two peaks. The values for $\alpha_M$ do agree well between peaks but, as expected are very different for the two ions of such different masses. 

\begin{table}
\centering
\bgroup
\def\arraystretch{1.3}
\begin{tabular}{c| c c | c c | c}
     & \multicolumn{2}{c|}{He$^+$} & \multicolumn{2}{c|}{Ne$^+$}& Units \\
     \hline
                & $E'$ & $A'_1$& $E'$ & $A'_1$\\ \hline
     $\sigma_1$ & 3.3 & 8.4  & 2.7 & 7.1 & 1 $\times10^{12}$ defects/cm$^{2}$ \\ 
     $\alpha_M$ & 0.0067& 0.0071 & 0.0976&0.1081& \textit{dimensionless} \\ 
\end{tabular}
\egroup
    \caption{
    (a) Initial defect density ($\sigma_1$) and defect yields ($\alpha_M$) calculated from fitting equation \ref{eq:ldmos2} to data as shown in Fig. \ref{fig:AMoS2Ratios} (b).}
    \vspace{0ex}
    \label{table:mos2results}
\end{table}

\section{Conclusion}
In this study, we explained the effects of both He$^+$ and Ne$^+$ irradiation at 30 keV on graphene and MoS$_2$ using models of their Raman spectra. For both materials, doses above 5$\times$10$^{15}$ cm$^{-2}$ and 5$\times$10$^{14}$ cm$^{-2}$ for the respective ion species have resulted in severe changes in the spectra relative to the starting material. For both material systems, this severe breakdown occurs at a Ne$^+$ dose which is between one and two orders of magnitude less than that of He$^+$. We believe this represents the first accurate experimental comparison of defect sizes produced in 2D materials by different noble gas ion probes.
The dose dependence of irradiation species and relationships to interdefect distance have been established for graphene. A clear comparison to the literature is also reported for MoS$_2$. In addition, we studied the effects of both primary ion species and secondary particles on the sizes of the defects produced in graphene. The role of substrate particles on defect production and size are also highlighted as a concern for nanofabrication methodologies due to the clear secondary atom effect. 
These results will allow a more informed and precise defect engineering of the investigated monolayer materials.  

\section{Acknowledgements}
We thank the staff at both the Photonics and the Advanced Microscopy Laboratories (AML), CRANN, Trinity College Dublin and at the National Laboratory of Solid State Microstructures, Nanjing University. We acknowledge support from the following grants: Science Foundation Ireland [grant numbers: 12/RC/2278, 11/PI/1105, 07/SK/I1220a, 15/IA/3131 and 08/CE/I1432]. 
\FloatBarrier
\clearpage

\bibliography{library}

\end{document}